# NATURAL RADIOACTIVITY MEASUREMENTS APPLIED TO THE DATING AND AUTHENTICATION OF EDIBLE MEAT

Michael S. Pravikoff[1], Philippe Hubert[1] and Hervé Guégan[2]

## ABSTRACT

The measurement of the ratio of $^{228}$Th vs $^{228}$Ra, two isotopes from the natural radioactivity, is an ancient, but not entirely reliable, method of dating biological materials. It has been applied for the dating of fishes [1] with mitigated results in many cases. It has been also experimented with mixed success in forensics cases [2,3] where the question of how long a person whose bones were found was dead. Here too, the precision of the results is not satisfactory. Other applications concern age dating oil and gas wastewaters spills.[4]

Our goal, while based on the same measurement of the ratio of the two radiosiotopes, is different. We aim at assessing the age of an animal at the date of its death, in particular for the animals intended for human consumption. The measurement technique relies on gamma spectrometry of the bones of $^{228}$Th and $^{228}$Ra stemming from ingested food and from the environment. A strong incentive to our research is fraudulent sales of meat with falsified identification. For instance, mutton sold as lamb, which is more expensive. Depending on the country of origin, this is complicated by varying legal definitions of the animal category, which renders controls for imported meat difficult. Preliminary measurements with retail samples of approximate identification validate the procedure. Further and more precise inquiries in a collaborative work with the French Ministry's Anti-Fraud Agency (equivalent of the U.S. Food and Drug Administration) are on their way.

## 1. PRINCIPLE OF DATING

It is based on the measurement of the imbalance between $^{228}$Ra (half-life of 5.7 years) and $^{228}$Th (half-life of 1.9 years). These two isotopes belong to the natural radioactive chain of $^{232}$Thorium (Figure 1). $^{232}$Th having a very long period of 14 billion years, is present almost everywhere on Earth. During its decay, it generates a series of daughter nuclei including $^{228}$Ra and $^{228}$Th. Because of its solubility in water, radium is therefore found almost everywhere in nature and in food at very different levels. After ingestion the radium dissolves in the blood, then, because of its chemical properties close to the calcium, it is fixed to a large extent in the bones. On the contrary, its descendant, $^{228}$Th

1 Centre d'Études Nucléaires de Bordeaux-Gradignan (CNRS/IN2P3 & Université de Bordeaux) - 19 chemin du Solarium, CS 10120, F-33175 Gradignan cedex, France

2 CENBG / PRISNA Prestations - 19 chemin du Solarium, CS 10120, F-33175 Gradignan Cedex, France

has a very low solubility: it is almost totally rejected by the body. Consequently, if a bone contains $^{228}$Th, its origin is only due to the decay of the mother nucleus, $^{228}$Ra, which is already fixed in the bone. The ratio of the levels (or activities) in $^{228}$Ra versus $^{228}$Th is time-dependent, hence the time elapsed between the birth of the animal and the date of measurement of the bone sample.

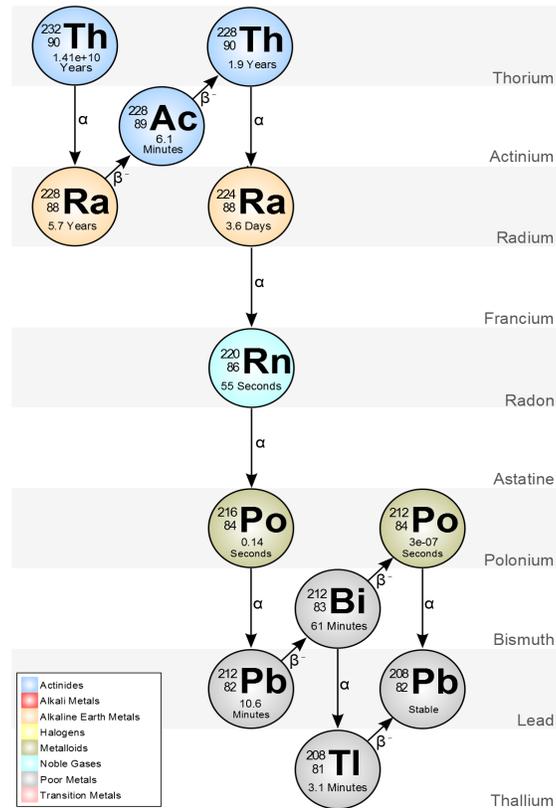

Figure 1

If the measurement of activities is done just after the slaughter of the animal, its age is directly obtained, otherwise the storage time is taken into account whether the sample has been frozen, canned or preserved. The dating is limited to about ten years, beyond which the ratio of activities $^{228}$Th and $^{228}$Ra tends to a constant, which explains the difficulties encountered in forensics for adult bones, not to mention the time elapsed between the death and the discovery of the remains.

On the other hand, this ratio evolves quickly and almost linearly during the first months and the inflection of the slope varies little until after a few years of age. Calculating the theoretical ratio according to the classical equations of radioactivity shows that it is very different for a young calf and and adult cow.

The theoretical calculations are based on the formulation of the ratio $A_2/A_1$ ($^{228}$Th / $^{228}$Ra) of the respective activities of the two isotopes:

$$\left(\frac{A_2}{A_1}\right)_t = \frac{(1 - e^{-\lambda_2 t}) - \left(\frac{\lambda_2}{\lambda_2 - \lambda_1}\right)(e^{-\lambda_1 t} - e^{-\lambda_2 t})}{(1 - e^{-\lambda_1 t})}$$

In the case of poultry whose life before slaughter last only a few months, it is easy to discriminate a chicken housed and raised in battery cages from top-quality free-range poultry.

## 2. PROTOCOL FOR THE PREPARATION OF SAMPLES AND MEASUREMENTS

All bone samples are reduced to ashes. After cleaning, each bone is heated gradually to 500° C and maintained at this temperature for 8 h. After cooling, the sample is crushed to a fine powder, which fills a 5 $cm^3$ polyethylene tube; This tube which is then placed in one of the two low-noise background high-purity gamma-ray Ge detectors.

These spectrometers are made with materials non-radioactive or with very weak radioactivity. They are of the "well" type and are installed at the PRISNA (Plate-forme Régionale Interdisciplinaire de Spectrométrie Nucléaire en Aquitaine) platform at the Centre d'Etudes Nucléaires de Bordeaux-Gradignan (CENBG) (see picture below). Shielded from natural external radioactivity by lead, with the innermost being a non-radioactive archeological lead dating from the end of the Roman Empire, additional boron-filled polyethylene plates surround the spectrometer to attenuate the neutron background noise.

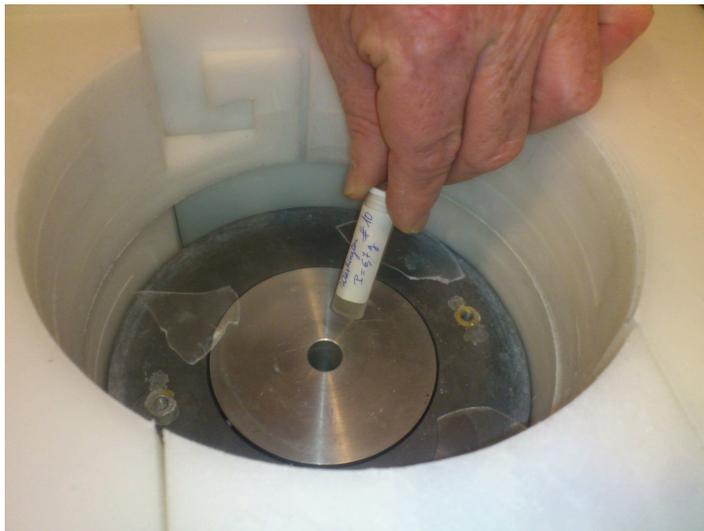

The results of the measurements are energy spectra in which the gamma peaks (or lines) are identified qualitatively and quantitatively. Each line, by its energy in particular, is characteristic of a given radioactive nucleus/isotope. The spectra provides therefore a direct signature of the gamma emitting radioisotopes content of the sample tested. This is the big advantage over other measurement methods (alpha- and beta-counting) which involve a chemical separation processes and destruction of the probe.

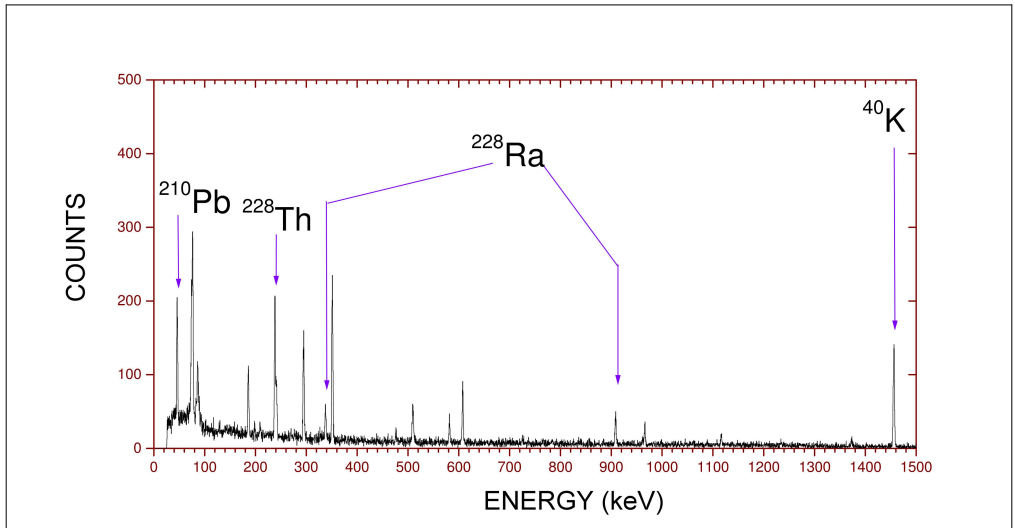

Figure 2

An example of the spectrum obtained is shown in Figure 2 in the case of a sheep bone. The gamma rays associated with $^{228}$Th (238 keV) and $^{228}$Ra (338 keV and 911 keV), used for dating, are clearly apparent. Note the presence of other natural radioactive isotopes such as $^{210}$Pb and $^{40}$K.

## 3. EXPERIMENTAL MEASUREMENTS

The level of radium in the muscles being very low, mostly at the limit of detection, dating is only possible for a bone sample.

### 3.1 POULTRY

Figure 3 shows the results of bone measurements of standard chickens, red label farm chickens and Bresse ones (the only kind having a protected designation of origin). The slaughter ages are 40-42 days, greater than 81 days and greater than 112 days resp. The points in hollow squares beyond 6 months correspond to chickens procured directly at the inhabitant. The full squares are the farm and Bresse chickens whose traceability was known; They were remeasured after a wait of about 5-6 months, which confirmed the change of the ratio as was expected.

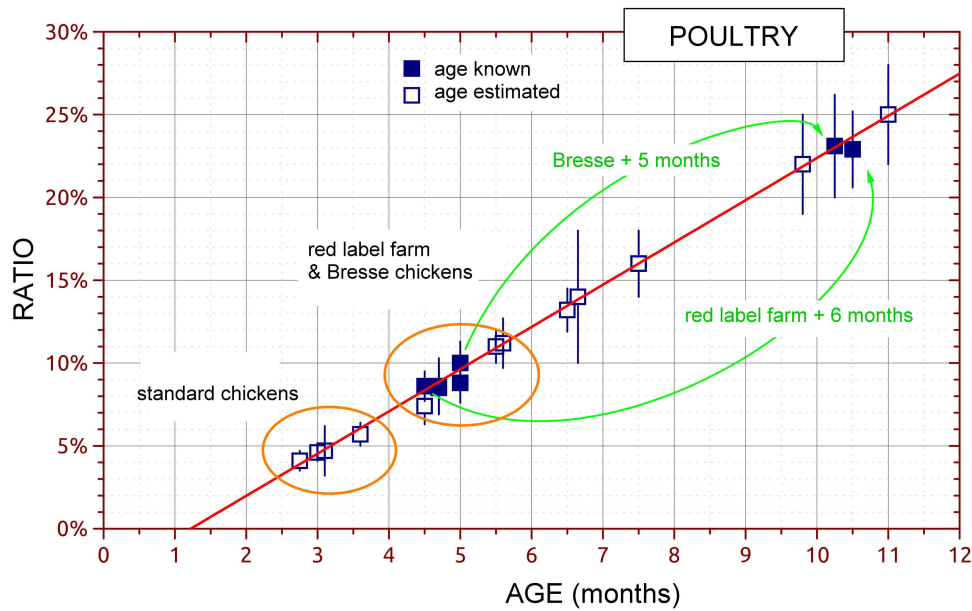

Figure 3

## 3.2 LAMBS

A larger time span was available for lamb samples. And for milk-fed lambs from New Zealand, comprehensive records of the animal were at hand. Results are shown in Figure 4. The two triangles are the frozen New Zealand lamb samples. They appear at an apparent age of 15 and 21 months respectively, because it includes the time elapsed since slaughter. The square at 25 months is that of a young sheep and no longer a lamb.

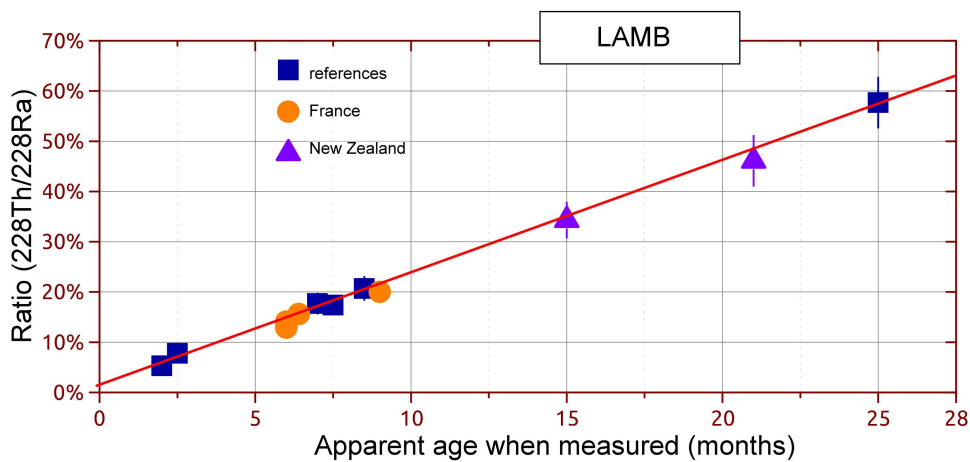

Figure 4

## 3.3 COWS

Measurements of calf and cow bones are shown in Figure 5. The fragmented information we had leaves in some cases a fairly large uncertainty about the actual age of the animals. However, there is a clear difference between calves and cows. As for cows, most of those sold under the name of "beef" are aged between 4 and 6 years if one trusts the curve. Superannuated cattle are much older as evidenced by the measurements.

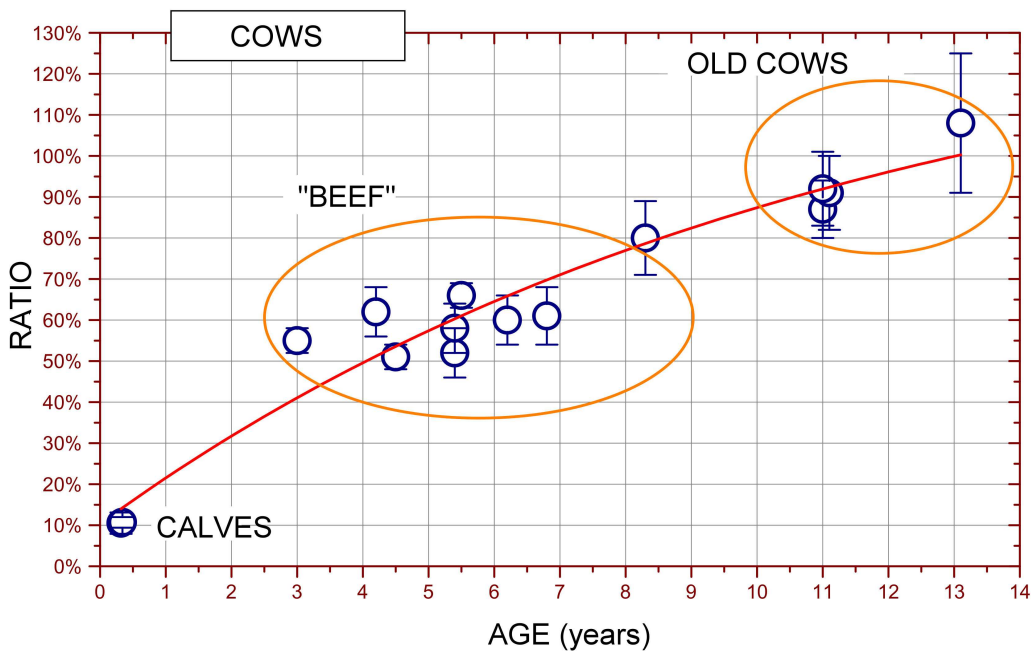

Figure 5

## 4. THEORETICAL CURVE, EXPERIMENTAL DATA AND CONCLUSION

Figure 6 shows the theoretical curve drawn according to the equation of §1 with experimental values for lamb and sheep bones plotted on the graph. The gamma spectrometry dating technique of the radioactivity contained in the bones of a dead animal leads to very satisfactory results which will be improved by new generations of low background gamma detectors placed in environments protecting them from interference due to natural radioactivity (such as PRISNA or any underground location or under a large rock mass (for example at the Modane Underground Laboratory in the Fréjus Road Tunnel ).

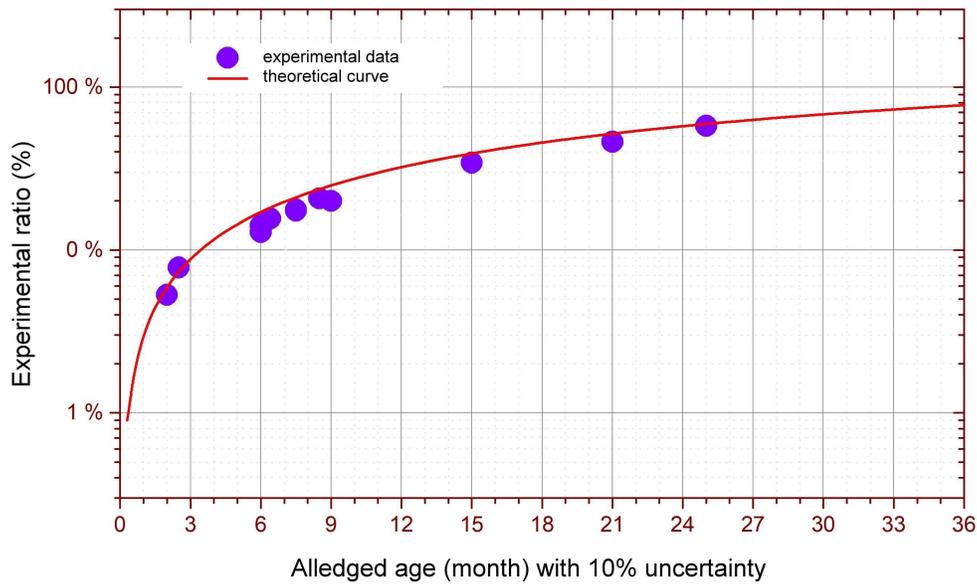

Figure 6